\documentclass{ws-ijmpc}

\usepackage{url}
\usepackage[%
  colorlinks=true,
  urlcolor=blue,
  linkcolor=blue,
  citecolor=blue
]{hyperref}
\hypersetup{linkcolor=red, citecolor=red, colorlinks=true}

\usepackage{mathtools,amsmath}
\usepackage{amsmath,amssymb}

\begin{document}

\markboth{Jaime L. C. da C. Filho and N. Crokidakis}
{Opinion dynamics under electoral shocks in competitive campaigns}

\catchline{}{}{}{}{}

\title{Opinion dynamics under electoral shocks in competitive campaigns}

\author{Jaime L. C. da C. Filho$^1$,
        Nuno Crokidakis$^2$
}

\address{
$^{1}$ Instituto Federal do Pará, Campus Castanhal, Castanhal/PA,  \hspace{1mm} Brazil\\
 jaime.filho@ifpa.edu.br\\
$^{2}$Instituto de F\'{\i}sica, \hspace{1mm} Universidade Federal Fluminense \hspace{1mm} Niter\'oi/RJ, \hspace{1mm} Brazil}

\maketitle

\begin{abstract}
We propose a computational framework for modeling opinion dynamics in electoral competitions that combines two realistic features: voter memory and exogenous shocks. The population is represented by a fully-connected network of agents, each holding a binary opinion that reflects support for one of two candidates. First, inspired by the classical voter model, we introduce a memory-dependent opinion update: each agent's probability of adopting a neighbor's stance depends on how many times they agreed with that neighbor in the agent's past $m$ states, promoting inertia and resistance to change. Second, we define an electoral shock as an abrupt external influence acting uniformly over all agents during a finite interval $[t_0, t_0+\Delta t]$, favoring one candidate by switching opinions with probability $p_s$, representing the impact of extraordinary events such as political scandals, impactful speeches, or sudden news.  We explore how the strength and duration of the shock, in conjunction with memory length, influence the transient and stationary properties of the model, as well as the candidates' advantage. Our findings reveal a rich dynamical behavior: memory slows down convergence and enhances system resilience, whereas shocks of sufficient intensity and duration can abruptly realign collective preferences, particularly when occurring close to the election date. Conversely, for long memory lengths or large election horizons, shock effects are dampened or delayed, depending on their timing. These results offer insights into why some sudden political events reshape electoral outcomes while others fade under strong individual inertia. Finally, a qualitative comparison with real electoral shocks reported in opinion polls illustrates how the model captures the competition between voter inertia and abrupt external events observed in actual elections.

\keywords{Opinion dynamics; Electoral shocks; Opinion inertia; Collective phenomena}
\end{abstract}

\section{\label{sec:intro}Introduction}

\qquad Opinion dynamics has emerged as a powerful interdisciplinary framework for analyzing how individual beliefs evolve through social interaction, offering mathematical and computational tools to explain collective behaviors \cite{castellano2009,parongama2014,csf_2023}. Its applications range from market trends \cite{martins2009,quanbo2020} and information diffusion \cite{guille2013,quattrociocchi2014,jaime2026} to electoral disputes \cite{costafilho1999,bernardes_epjb,travieso2006}, where elections are modeled as large-scale decision-making processes shaped by peer influence and external factors \cite{fortunato2007,fernandezgarcia2014,sirbu2013,sirbu2016,sirbu2023}. Through agent-based and stochastic approaches, researchers can explore how micro-level interactions lead to macro-level outcomes such as consensus \cite{clifford1973,ligget1985,sznajd2000,jstat2011,jstat2016}, polarization \cite{deffuant2000,hegselmann2002,ijmpc_forgerini2024}, or abrupt opinion shifts \cite{michard2005}, systematically testing scenarios, for instance, involving campaign strategies \cite{gionis2013,javarone2014} and misinformation propagation \cite{acemuglo2011,shin2018}, thereby providing valuable insights into both predictive outcomes and the mechanisms underlying democratic stability and volatility \cite{mainwaring2007}.

In recent decades, several statistical physics approaches have been proposed to investigate electoral dynamics. Early applications of the Sznajd model reproduced empirical features of vote distributions in Brazilian elections, both on regular lattices \cite{bernardes_ijmpc} and on complex networks such as the Barab\'asi–Albert topology \cite{bernardes_epjb}. Extensions of these models have explained the scaling behavior in the observed power-law patterns of vote distributions \cite{nuno_celia_angelo}. More recently, opinion dynamics models were tailored to bipolar political systems, showing how global effects such as government rating or propaganda can influence individual decision-making beyond local interactions \cite{caruso}. Other works have explored abstention and its critical impact on electoral outcomes \cite{serge_abstention}, as well as the role of campaign expenditure in shaping vote distributions, revealing diseconomies of scale in electoral processes \cite{melo}. A recent work has shown that kinetic exchange opinion models may undergo an order–disorder–order transition as the noise parameter varies, with the statistical behavior of the winning margin scaling differently across these phases, thereby providing a bridge between model predictions and empirical electoral data \cite{biswas2025}.

Alongside these contributions, sociophysics models by Galam and collaborators have directly tackled high-profile elections. Galam proposed a framework to explain the unexpected victory of Donald Trump in 2016 by highlighting the role of inflexible minorities and hidden prejudices \cite{galam_trump1}, later extending the analysis to the 2020 U.S. election \cite{galam_trump2}. In parallel, the same theoretical tools were applied to the Brazilian context, where the surprising outcome of the 2018 Bolsonaro victory was analyzed through the lens of contagion-like opinion models, including the role of populist attitudes \cite{nuno_galam}. These studies emphasize how nonlinear mechanisms and abrupt shifts can redefine expected results, complementing the empirical-driven approaches of earlier works.

An electoral shock is an unexpected event or disruption - such as a political scandal, a sudden economic crisis, or a major social movement - that emerges close to an election and substantially reshapes voter perceptions and behavior. Since electoral outcomes often rest on fragile balances of opinion, even minor shocks can tip preferences, heighten polarization, or mobilize previously undecided voters \cite{fieldhouse}. Such dynamics may ultimately alter the result of an election, benefiting candidates who skillfully adapt to the new scenario while undermining those who become linked to the negative perceptions triggered by the shock.

Some researches had already analyzed analogous phenomena. Fortunato \cite{fortunato2005} used damage spreading processes \cite{stanley1987,grassberger1995} to investigate, in Krause-Hegselmann model, how much a sudden perturbation (interpreted as electoral shock) would alter opinions of the agents of a system. The study demonstrates that the impact of an initial perturbation may vary from negligible to complete, depending on the confidence parameter $\epsilon$. A major contribution of the work is the identification of a phase in which the entire system becomes highly sensitive to the initial shock, implying that the collective dynamics can be determined by a single local alteration.

Fortunato and Stauffer \cite{fortunato2006} also used damage spreading, but in the Sznajd model, to simulate the reactions of individuals to extreme events by altering the opinion of a single agent and examining how a consensus model propagates this perturbation throughout the community. The simulations reveal that the system is highly robust to perturbations once consensus is nearly reached, with late shocks exerting little to no influence on the final outcome. In contrast, early shocks can decisively alter the trajectory of the dynamics, potentially reversing the emerging majority opinion. The effectiveness of such shocks is strongly dependent on their intensity: below a critical threshold they have negligible impact, while above it they can completely determine the final consensus. Overall, the results indicate that the timing and strength of external events are crucial in shaping collective outcomes under the Sznajd dynamics.

Another line of research has highlighted the role of opinion inertia in shaping collective dynamics. The inclusion of inertial mechanisms modifies classical models, often leading to qualitatively new behaviors such as additional phase transitions or altered coarsening properties. For instance, inertia in the majority-vote model can induce discontinuous transitions and hysteresis effects \cite{harunari,chen}. In the voter model framework, persistent states and dynamically induced zealots generate surface-tension-like effects, bridging voter-like and Ising-like coarsening regimes \cite{latoski}. Empirical-inspired coordination games also reveal that individuals frequently follow a local majority rule tempered by inertia toward previous choices \cite{gaisbauer}. More recently, cognitive inertia has been investigated in the context of activity-driven models of social media, showing its impact on polarization and consensus through mechanisms akin to pitchfork bifurcations \cite{pranesh}. These studies support the relevance of memory and persistence effects, reinforcing the motivation for our model that explicitly incorporates individual memory as a stabilizing mechanism against abrupt electoral shocks.

Within this whole context, our contribution is twofold: we propose a memory-dependent voter-like model, where the probability of imitation depends on past opinion alignment, and we incorporate the effect of an electoral shock, a finite-time external perturbation representing sudden exogenous events such as scandals or impactful speeches. This combination allows us to explore how persistence and abrupt shocks jointly shape consensus and realignments in electoral competitions. Section 2 details the model and simulation framework, while Section 3 reports our main results for both transient dynamics and stationary states. Section 4 concludes with a summary of the findings and outlines possible directions for future extensions.


\section{Model}
\label{model}

\qquad Our model for studying the influence of an electoral shock on elections builds upon the original Voter model. In this framework, each agent holds a binary opinion variable representing the preference between two candidates, denoted as A and B. To better capture the dynamics of sudden and collective shifts in electoral preferences, we introduce two key modifications to the standard Voter model.

The first modification introduces the electoral shock, modeled as an external field acting uniformly on all voters. Within the interval $[t_0, t_0+\Delta t]$, this field induces individuals to adopt opinion $+1$ (candidate A) with probability $p_s$. Accordingly, voters holding opinion $-1$ may switch to $+1$ under its influence. This mechanism is designed to represent extraordinary real-world events, such as political scandals, impactful speeches, or breaking news, that can abruptly realign voter preferences.

The second modification introduces memory, allowing each voter to retain a record of their last $m$ choices. The memory of individual $i$ at time $t$ is defined as
$$
M_i(t)=[O_i(t-m), O_i(t-m+1), \dots, O_i(t)],
$$
where $O_i(t)$ denotes the opinion of voter $i$ at time $t$. The probability of adopting a neighbor's opinion then depends on how frequently that opinion has appeared in the voter's memory, thereby introducing persistence effects that may either slow down or accelerate opinion shifts depending on past convictions.

The opinion update dynamics largely follow the original Voter model. At each step, an individual  $i$ is randomly selected together with a randomly chosen neighbor $j$. The state of $i$ is then updated (i.e., we take $O_i=O_j$) according to
\begin{equation}
p_i=\frac{n_{ij}}{m},
\label{eq1}
\end{equation}
\noindent
where $n_{ij}$ is the number of times within its memory that voter $i$ has previously held the same opinion as voter $j$. For most of the simulation, only this memory-based interaction rule applies. During the interval $[t_0, t_0+\Delta t]$, however, the influence of the external field is also considered, introducing a temporary yet potentially decisive bias into the dynamics. When the shock occurs, however, the influence of the external field is also considered, introducing a temporary yet potentially decisive bias into the dynamics.

We consider a fully connected population of $N$ voters, where each individual may interact with any other. The initial fraction of supporters of candidate A is denoted by  $f_{A,0}$. In the absence of an electoral shock, candidate $A \,(O=+1)$ has virtually no chance of defeating candidate $B\,(O=-1)$. When the shock occurs, however, the induced alignment toward $+1$ can substantially increase candidate $A$'s competitiveness, even under initially unfavorable conditions.

As it is standard in agent-based simulations, in a population of $N$ voters one unit of simulation time corresponds to $N$ possible interactions, referred to as one Monte Carlo step (MCS). During the simulations, we monitored several observables, including the fraction of supporters of candidate $A$, the victory probability of candidate $A$, and the switching probabilities toward candidates $A$ and $B$, to evaluate their time evolution. To ensure statistical robustness, final results were obtained by averaging over multiple independent simulation runs.

By combining the effects of an external opinion shock with individual memory, our approach accounts simultaneously for the short-term volatility and the long-term inertia observed in real electoral processes \cite{fieldhouse,green}, thus providing a richer framework to understand how sudden events may reshape competitive political scenarios.


\section{Results and discussions}
\label{results}

\qquad Unless otherwise stated, our results were obtained for a population of $N=1000$ voters, with an initial support fraction $f_{A,0}=0.1$, and averaged over $r=500$ independent realizations. Additional tests with $N \in \{10, 100, 1000, 10000\}$ and $r \in \{500, 1000, 2000\}$ confirmed that only the smallest system size produced noticeable deviations, while larger values of $N$ and $r$ yielded results indistinguishable from those obtained with $N=1000$ and $r=500$. We therefore adopted these values as a balance between statistical accuracy and computational efficiency. The choice of $f_{A,0}=0.1$ represents a highly unfavorable initial condition for candidate A.

In addition, it was necessary to specify the initial memory configuration of the voters. To this end, the first $m-1$ memory slots of each individual were randomly assigned, with equal probability of being $+1$ or $-1$, while the most recent slot always corresponded to the individual's current opinion. This initialization does not affect the subsequent impact of the electoral shock on the dynamics; its influence is restricted to the very early stages of the process


\subsection{Impact of memory and electoral shock on stationary states}

\qquad To better understand the dynamics underlying the electoral dispute, we first examine the two mechanisms of interest—individual memory and electoral shock, separately. This step allows us to disentangle their independent contributions before analyzing their combined effects. By isolating each factor, we can determine how memory alone influences the persistence or reversal of opinions, and how a transient shock alters the trajectory of the electoral outcome.

Accordingly, we present the results in two stages. First, we focus on the role of memory, showing how the retention of past opinions shapes the long-term stability of the system. Next, we turn to the case of an electoral shock in the absence of memory, highlighting the temporary but significant disturbances it generates. This separate discussion provides a clear foundation for the subsequent analysis, where both mechanisms are considered simultaneously to capture their interaction effects.

Finally, we address the joint scenario, which enables us to observe how the interplay between memory and shocks modifies the overall opinion dynamics. This integrated perspective offers a more realistic representation of electoral processes, revealing whether the coexistence of memory and external perturbations reinforces, mitigates, or transforms the patterns observed in the separate analyses.


\subsubsection{Memory effects on opinion dynamics}

\begin{figure}[ht]
\centering
\includegraphics[width=0.7\linewidth]{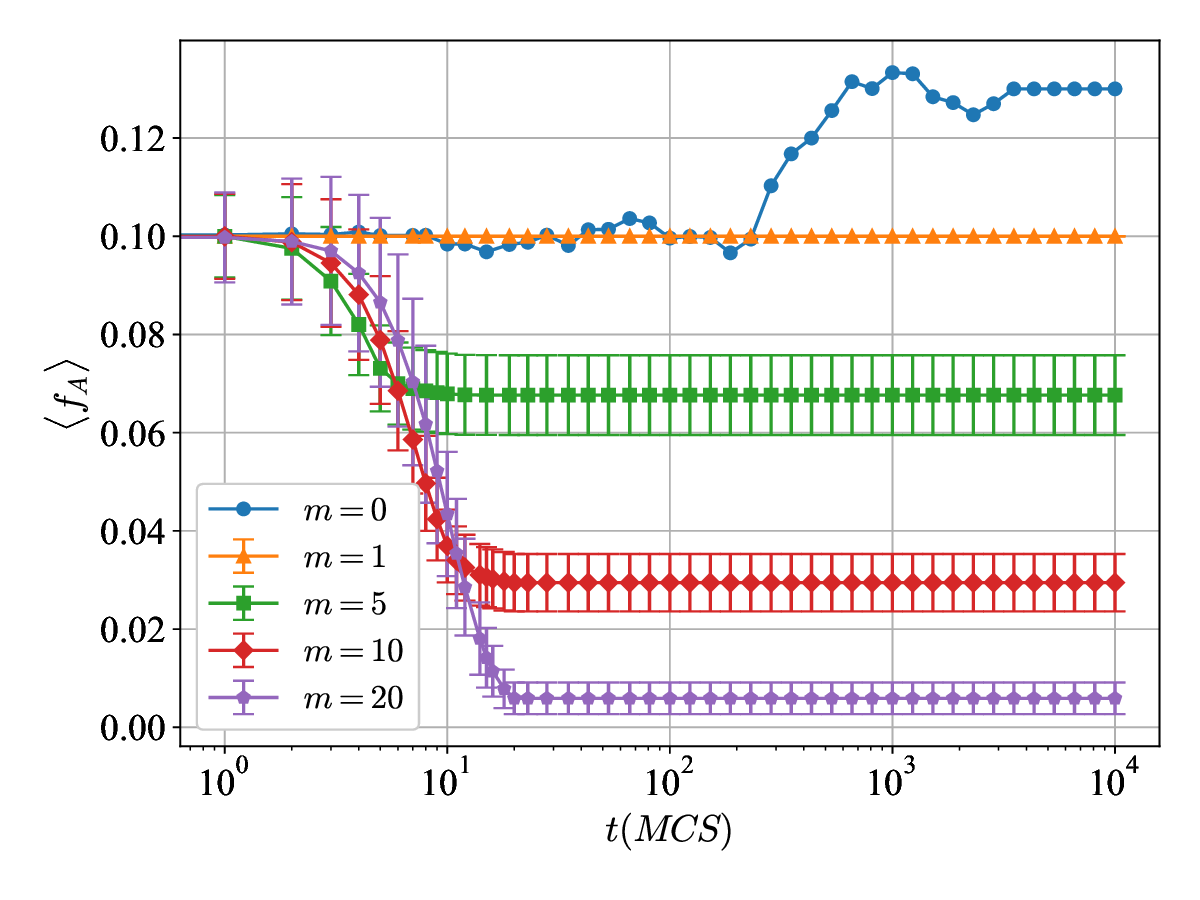}
\caption{Time evolution of the mean fraction of electors supporting candidate $A$ (main panel), for typical values of the memory size $m$. To analyze the impact of memory on the population, we did not considered the electoral shock in such simulations. The population size is $N=1000$, the initial fraction of $A$ opinions is $f_{A,0}=0.1$, and results are averaged over $500$ independent simulations.}
\label{fig1}
\end{figure}

\qquad Figure \ref{fig1} shows the time evolution of the mean fraction of voters supporting candidate $A$, where the ensemble average $\langle f_A\rangle$ was computed as an average over $500$ independent simulations. The case $m=0$ corresponds to the memoryless limit, i.e., the original Voter model. In this scenario, it is well known that each simulation evolves toward an absorbing state, either $f_A(t \to \infty)=1$ or $f_A(t \to \infty)=0$, with the probability of reaching consensus at $+1$ equal to the initial fraction of agents holding that opinion. In other words, the fraction of runs ending in consensus at $+1$ is directly given by the initial condition $f_{A,0}$. Consequently, for $m=0$, as illustrated in Figure \ref{fig1}, the long-time solution satisfies $\langle f_A(t \gg 0)\rangle \approx f_{A,0}$. The characteristic time to reach consensus—i.e., to attain a constant value of $\langle f_A\rangle$, is of order $10^3$ Monte Carlo steps for $N=1000$.

As shown in Figure \ref{fig1}, the dynamics of $\langle f_A\rangle$ exhibit marked differences between the cases $m=0$ and $m \neq 0$. In the former, $\langle f_A\rangle$ rapidly reaches a constant value. For $m=1$, no variation is observed, since individual memories prevent opinion changes. For $m>1$, $\langle f_A\rangle$ decreases quickly from its initial value but soon stabilizes at a constant level, which becomes lower as $m$ increases.

This markedly different behavior can be explained by analyzing the average switching probabilities toward candidate $A$, $\langle p_A\rangle$, and toward candidate $B$, $\langle p_B\rangle$, where the averages are taken over all simulations. These probabilities can be factorized as
\begin{equation}
p_{A(B)} = p_{A(B)}^{\text{int}} \cdot p_{A(B)}^{\text{mem}},
\end{equation}
\noindent
where $p_{A(B)}^{\text{int}}$ and $p_{A(B)}^{\text{mem}}$ denote, respectively, the probabilities of switching to $A$ ($B$) due to voter interactions and due to memory effects. Since these contributions are independent, the averages factorize as $\langle p_{A(B)} \rangle = \langle p_{A(B)}^{\text{int}} \rangle \cdot \langle p_{A(B)}^{\text{mem}} \rangle$. For the interaction-related probability, we obtain
\begin{equation}
\langle p_{A(B)}^{\text{int}} \rangle = p_{A(B)}^{\text{int}} = \frac{n_{A(B)}(N-n_{A(B)})}{N(N-1)},
\end{equation}
\noindent
where $n_{A(B)}$ is the number of voters supporting candidate $A$ ($B$); this probability is identical for all voters of the opposite candidate. The memory-related probability, in contrast, depends on the individual histories of voters. In this case, we employ Eq. (\ref{eq1}) for each voter of $B$ ($A$), and then take an average first over all such voters and subsequently over all simulations. Thus, $\langle p_{A(B)}^{\text{mem}}\rangle$ represents a double average.

\begin{figure}
\centering
\includegraphics[width=0.7\linewidth]{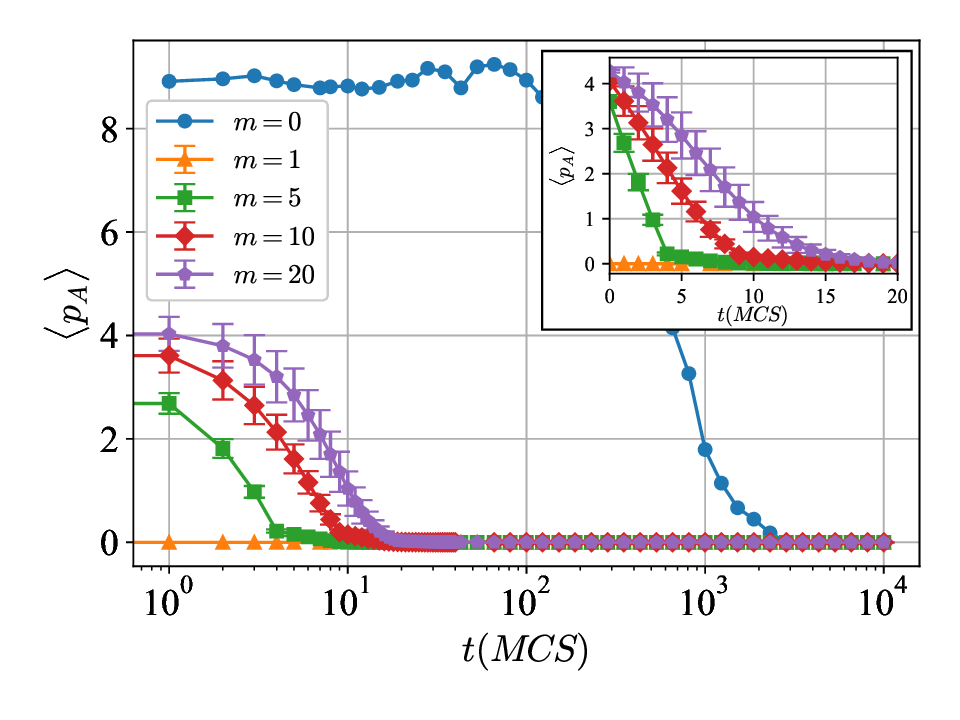}
\includegraphics[width=0.7\linewidth]{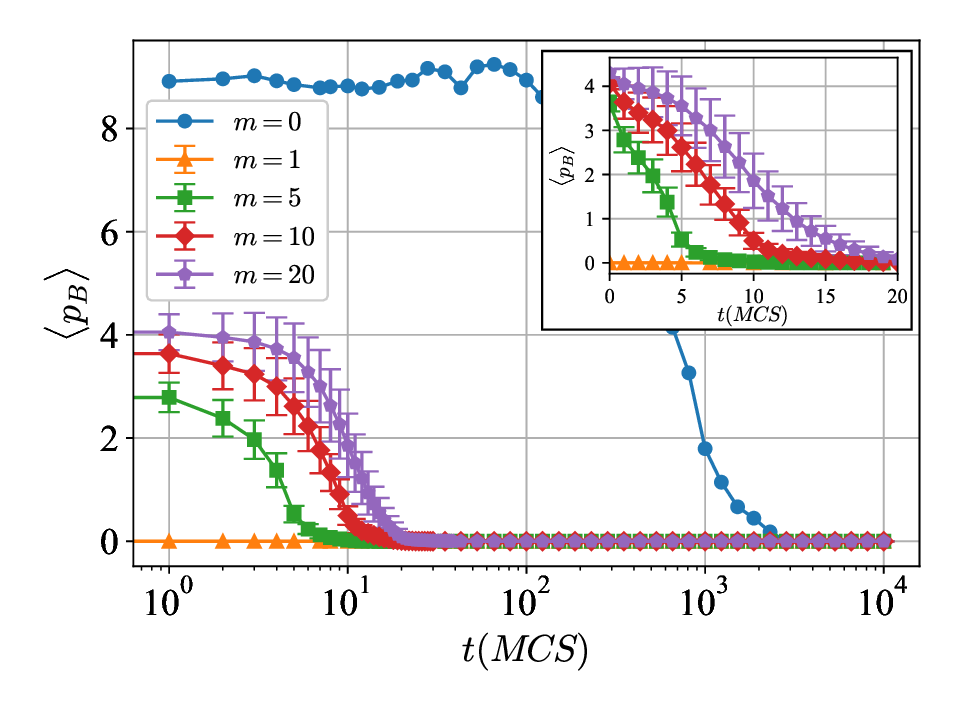}
\caption{Time evolution of the average switching probability toward candidate $A$ ($\langle p_A\rangle$, top panel) and toward candidate $B$ ($\langle p_B\rangle$, bottom panel) for different memory sizes $m$. Insets highlight the early stage of the dynamics for the cases with $m\neq 0$. Parameters: $N=1000, f_{A,0}=0.1$ and results are averaged over $500$ independent simulations.}
\label{fig2}
\end{figure}

Figure \ref{fig2} shows the time evolution of $\langle p_A \rangle$ (top panel) and $\langle p_B \rangle$ (bottom panel) for several values of $m$. In the memoryless case ($m=0$), we have $p_{A(B)} = p_{A(B)}^{\text{int}}$; here, $\langle p_{A(B)}\rangle$ remains finite up to times of order $10^3$ (for $N=1000$), after which the dynamics freezes. As expected, for $m=1$ the switching probabilities vanish identically, i.e., $\langle p_{A(B)}\rangle = 0$ throughout the entire evolution. For $m>1$, $\langle p_{A(B)}\rangle$ persists only up to times of order $10^1$ (for $N=1000$), which explains the faster freezing of $\langle f_{A(B)}\rangle$ compared to the $m=0$ case. Thus, the introduction of memory accelerates the freezing of the dynamics because $\langle p_{A(B)}^{\text{mem}}\rangle$ rapidly decays to zero when $m>1$. These results highlight the inertial role of memory: by suppressing switching probabilities, memory enforces persistence of past choices and drives the system more rapidly into frozen configurations.

As shown by comparing the top and bottom insets of Figure \ref{fig2}, the inequality $\langle p_B\rangle \geq \langle p_A\rangle$ holds at all times prior to the vanishing of the curves. This asymmetry arises from the mismatch between the initial memory and opinion configurations: while the former was initialized with equal preferences for candidates $A$ and $B$, the latter exhibits a strong bias toward candidate $B$. As the dynamics evolve, this initial bias progressively imprints itself on the agents’ memories. Consequently, we have $\langle p_B^{\text{int}}\rangle > \langle p_A^{\text{int}}\rangle$ together with $\langle p_B^{\text{mem}}\rangle \approx \langle p_A^{\text{mem}}\rangle$ at early times. This explains the initial decrease of $\langle f_A\rangle$ observed in Figure \ref{fig1}.


\subsubsection{Electoral shock effects on opinion dynamics}

\begin{figure}
\centering
\includegraphics[width=0.7\linewidth]{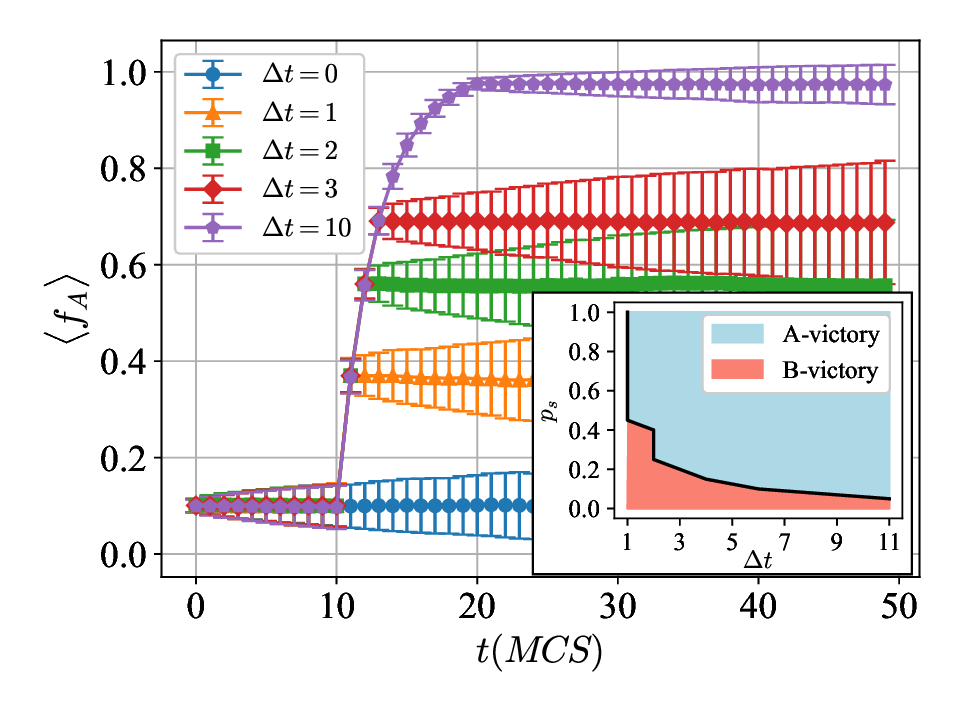}
\caption{Results for the memoryless version of the model ($m=0$). The main panel shows the time evolution of the mean fraction of electors supporting candidate $A$, $\langle f_A\rangle$, under an electoral shock applied at $t_0=10$ with $p_s=0.3$, for several values of $\Delta t$. The inset displays the $p_s \times \Delta t$ plane with regions of victory for candidates $A$ and $B$, obtained in the stationary regime ($t \gg t_0$). Parameters: $N=1000$, $f_{A,0}=0.1$.}
\label{fig3}
\end{figure}

\qquad In the main panel of Figure \ref{fig3}, we show the time evolution of $\langle f_A\rangle$ under the effect of an electoral shock applied at $t_0=10$ with $p_s=0.3$, for several values of $\Delta t$. In all cases, the average dynamics of $\langle f_A\rangle$ are affected only during the shock interval. Before $t_0$, the trajectories coincide at $\langle f_A\rangle = f_{A,0}$. After $t_0+\Delta t$, however, each trajectory stabilizes at a distinct constant value of $\langle f_A\rangle$, which increases monotonically with larger $\Delta t$. Notably, for $p_s=0.3$, shocks lasting $\Delta t \geq 2$ are already sufficient, on average, to reverse the electoral outcome in favor of candidate $A$.

To provide a more complete picture of shock events throughout the dynamics, the inset of Figure \ref{fig3} shows the $p_s \times \Delta t$ plane divided into two regions: the red area corresponds to values of $(p_s,\Delta t)$ where the electoral shock does not reverse the election outcome, while the light blue area represents the opposite case. This phase diagram was obtained in the stationary regime ($t \gg t_0$), and the black line indicates the threshold where the average outcome switches from $\langle f_A\rangle < 0.5$ to $\langle f_A\rangle > 0.5$. A small region of non-effectiveness is observed at low values of both $p_s$ and $\Delta t$, extending toward larger $\Delta t$ when $p_s$ remains small. Conversely, the largest portion of the parameter space corresponds to effective shocks. This pattern is expected, since the effectiveness of an electoral shock scales with both $p_s$ and $\Delta t$, as both parameters act in favor of candidate $A$. It is worth noting that when $p_s \approx 0.45$, candidate $A$ wins the election regardless of the shock duration $\Delta t$. These results confirm that the impact of electoral shocks depends critically on both their intensity ($p_s$ and duration ($\Delta t$). In the next subsection, we examine how this mechanism interacts with individual memory, revealing whether the presence of persistence amplifies, mitigates, or transforms the effectiveness of shocks in shaping electoral outcomes.

\subsubsection{Interplay between memory and electoral shocks}

\begin{figure}[ht]
\centering
\includegraphics[width=0.7\linewidth]{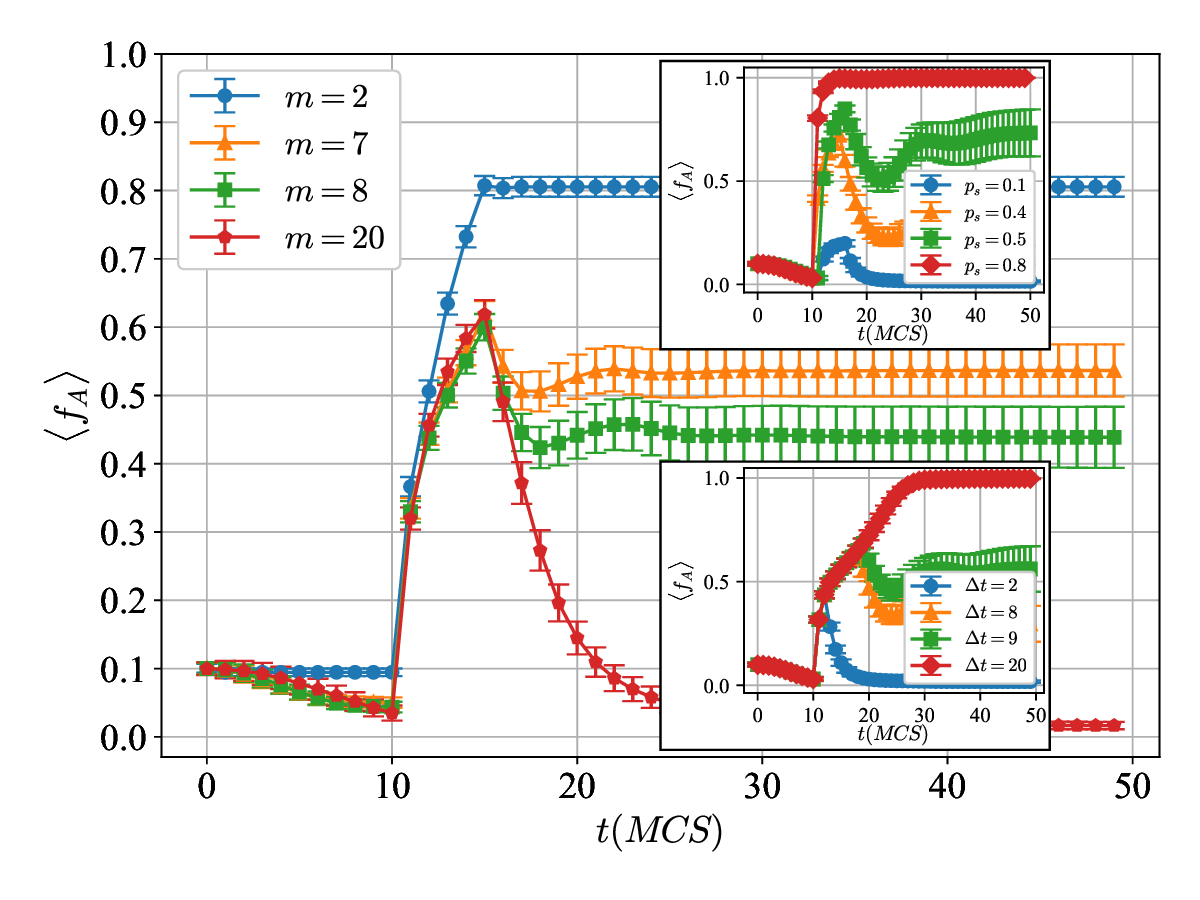}
\caption{Time evolution of $\langle f_A\rangle$ under an electoral shock applied at $t_0=10$. Main panel: $p_s=0.3$, $\Delta t=5$, for several values of $m$. Top inset: $\Delta t=5$, $m=15$, for several values of $p_s$. Bottom inset: $p_s=0.3$, $m=15$, for several values of $\Delta t$. Parameters: $N=1000$, $f_{A,0}=0.1$, results averaged over $r=500$ independent runs.}
\label{fig4}
\end{figure}

\qquad We now turn to the joint effect of both mechanisms. Figure \ref{fig4} shows the time evolution of $\langle f_A\rangle$ when an electoral shock is applied at $t_0=10$. As expected, the results display the characteristic pattern of the shock acting as a transient phenomenon, with its influence confined to the interval of application. The figure also allows us to assess how memory modifies the system's response to the shock, revealing whether persistence amplifies or mitigates its effectiveness in altering the long-term outcome.

The main panel of Figure \ref{fig4} presents results for $p_s=0.3$ and $\Delta t=5$, considering different values of $m$. For shorter memory lengths ($m \leq 7$), candidate $A$ surpasses candidate $B$ around $t \approx 13$ ($\langle f_A\rangle > 0.5$), establishing a stable lead that persists until the end of the simulation. In contrast, for longer memory lengths ($m > 7$), the system exhibits only a transient response: after peaking at $t=t_0+\Delta t$, $\langle f_A\rangle$ gradually returns to favor candidate $B$. In this regime, candidate $A$ experiences increased support only within a narrow time window immediately before the shock ends. For $m=20$ (and in fact for all $m \geq 15$, omitted here for clarity), $\langle f_A\rangle$ almost completely recovers its pre-shock value. This difference in behavior reflects the stronger capacity of long memories to retain past opinions, which counteracts the influence of the temporary external field and accelerates the return to the pre-shock state. These findings demonstrate that memory can strongly buffer the impact of electoral shocks, to the point of almost neutralizing their long-term effects when past opinions are retained over sufficiently long horizons.

The top inset of Figure \ref{fig4} further analyzes the time evolution of $\langle f_A \rangle$ for $\Delta t=5$ and $m=15$, considering different values of $p_s$, which represents the strength of the electoral shock. The results indicate that for $p_s \geq 0.5$, candidate $A$ gains sufficient momentum for the race to shift decisively in his favor. Conversely, for $p_s < 0.5$, the effect remains transient, with no lasting advantage. The bottom inset of Figure \ref{fig4} shows the time evolution of $\langle f_A \rangle$ for $p_s=0.3$ and $m=15$, considering different shock durations $\Delta t$. In this case, $\Delta t=9$ marks the threshold beyond which candidate $A$ secures victory in the electoral race after the shock.

\begin{figure}[ht]
\centering
\includegraphics[width=0.7\linewidth]{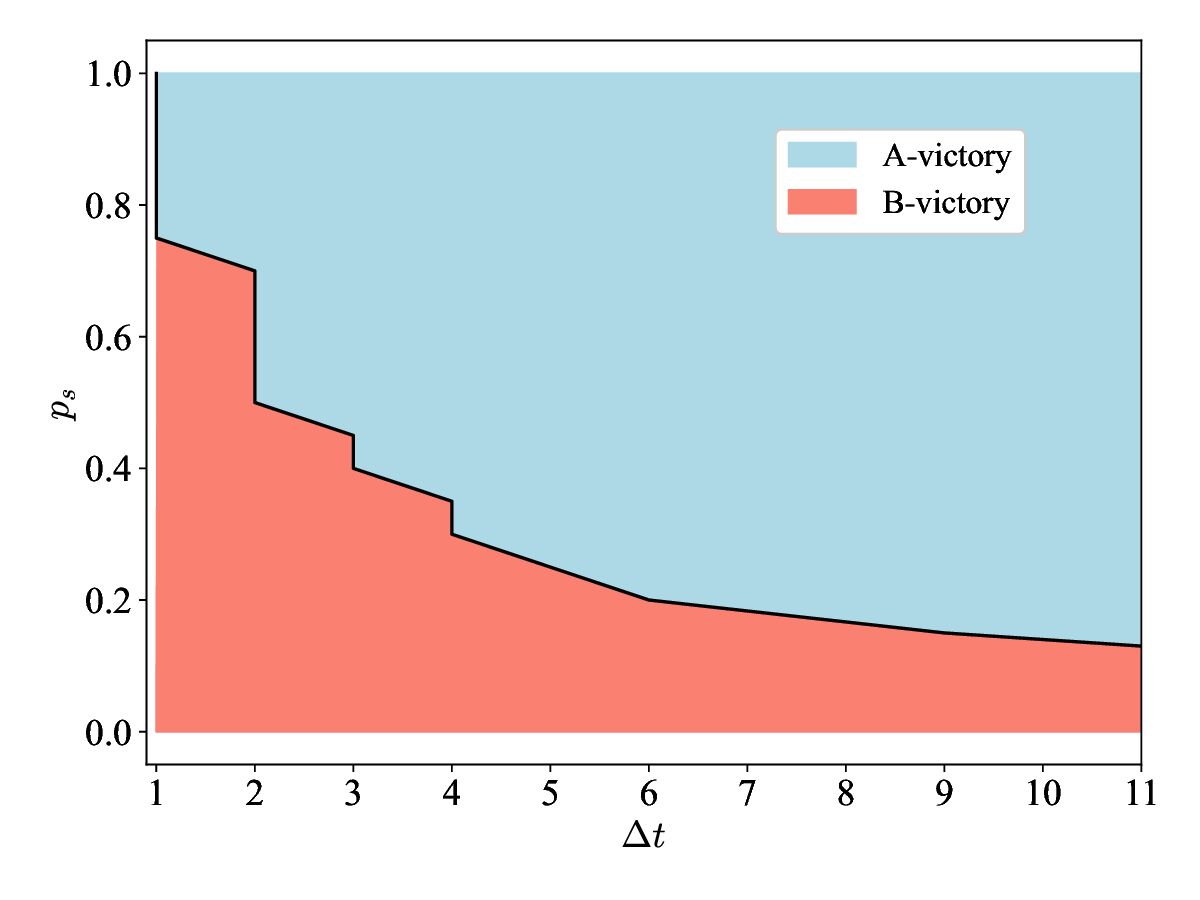}
\caption{Victory regions of candidates $A$ and $B$ in the $p_s \times \Delta t$ plane, obtained in the stationary regime ($t \gg t_0$) for $m=5$. Parameters: $N=1000$, $f_{A,0}=0.1$, results averaged over $r=500$ independent runs.}
\label{fig5}
\end{figure}

Figure \ref{fig5} shows the regions of victory for candidates $A$ and $B$ in the $p_s \times \Delta t$ plane for $t \gg t_0$, analogous to the inset of Figure \ref{fig3}, but now with $m=5$. Compared to that earlier case, a larger victory region for candidate $B$ is observed, and this region further expands as $m$ increases (not shown). In this scenario, only for $p_s \approx 0.75$ does candidate A secure victory independently of the shock duration $\Delta t$. Overall, these results demonstrate that longer memory lengths systematically reduce the effectiveness of electoral shocks, enlarging the victory region of candidate $B$ and requiring increasingly strong shocks for candidate $A$ to prevail.


\subsection{Short-time impact of memory and shocks: Favorable scenarios for candidate $A$}

\qquad In the previous analysis we examined the overall effects of memory and electoral shocks on the dynamics, focusing on the long-time behavior ($t \gg 0$) and the resulting stationary states. However, this time scale does not necessarily correspond to that of a real electoral competition. Moreover, the long-time analysis does not provide information about the short-time behavior immediately after the shock ($t \gtrsim t_0$). To address this, we introduce the variable $\tau$, defined as the elapsed time after $t_0$, and assume that $\tau$ represents the hypothetical election day (\textit{election horizon}). This perspective allows us to evaluate how memory and shocks influence electoral outcomes on shorter horizons, providing insight into the conditions under which candidate $A$ may achieve an advantage by the hypothetical election day.

\begin{figure}
\centering
\includegraphics[width=0.49\linewidth]{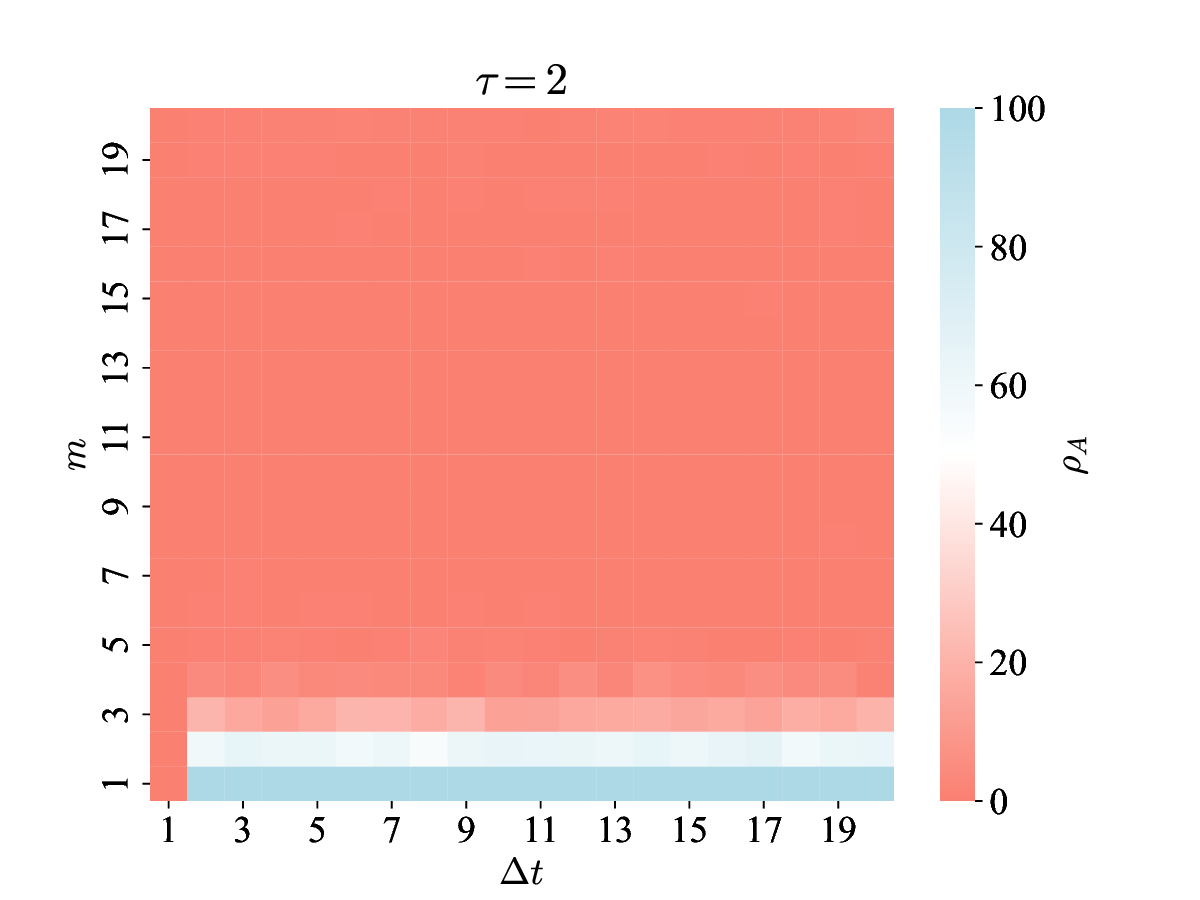}
\includegraphics[width=0.49\linewidth]{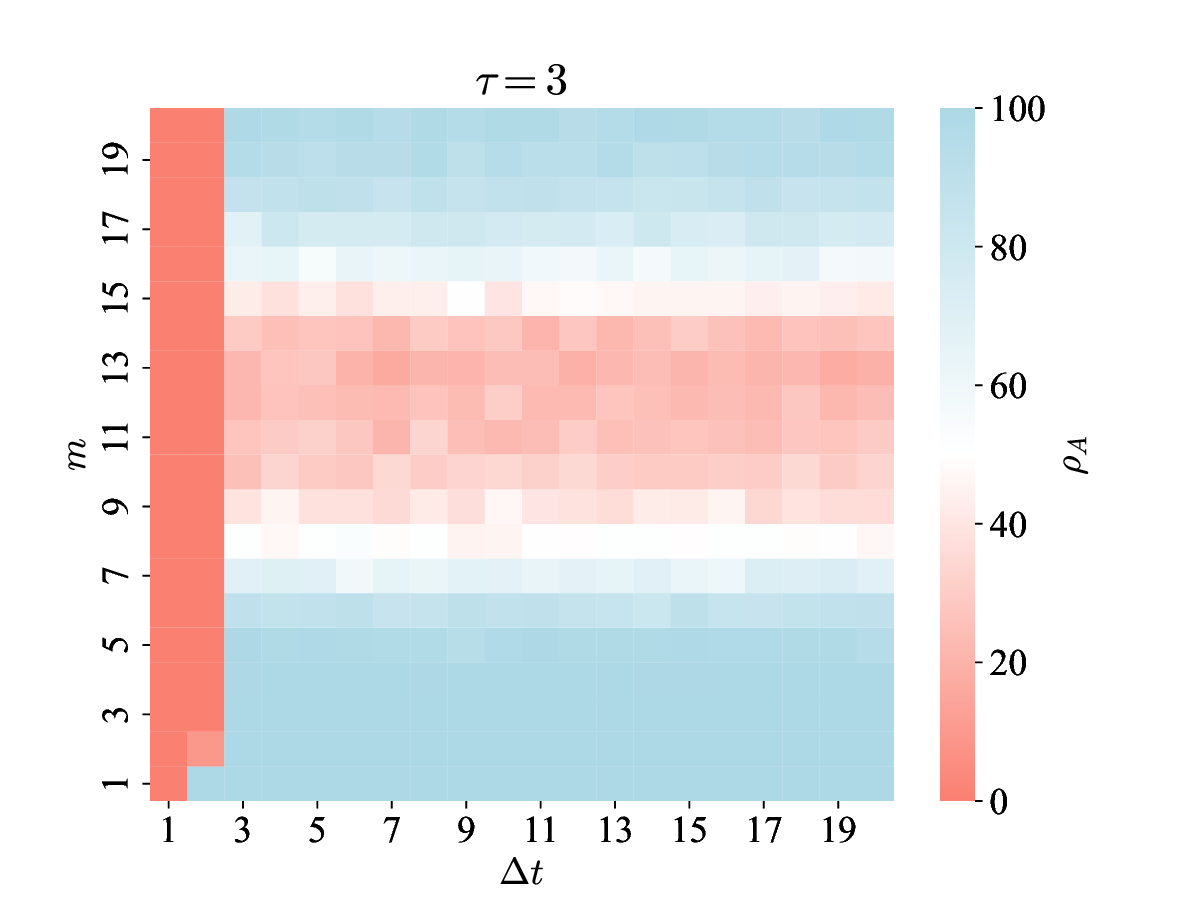}
\includegraphics[width=0.49\linewidth]{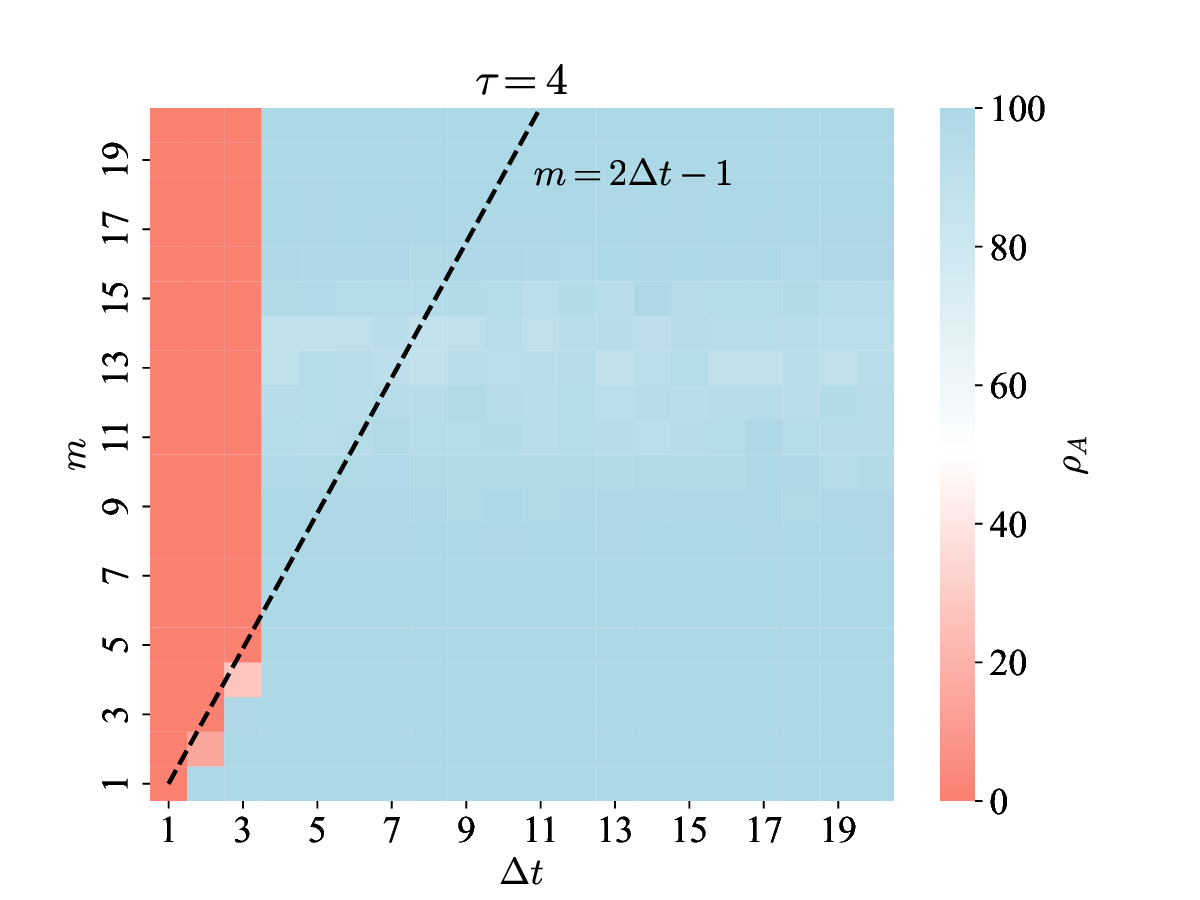}
\includegraphics[width=0.49\linewidth]{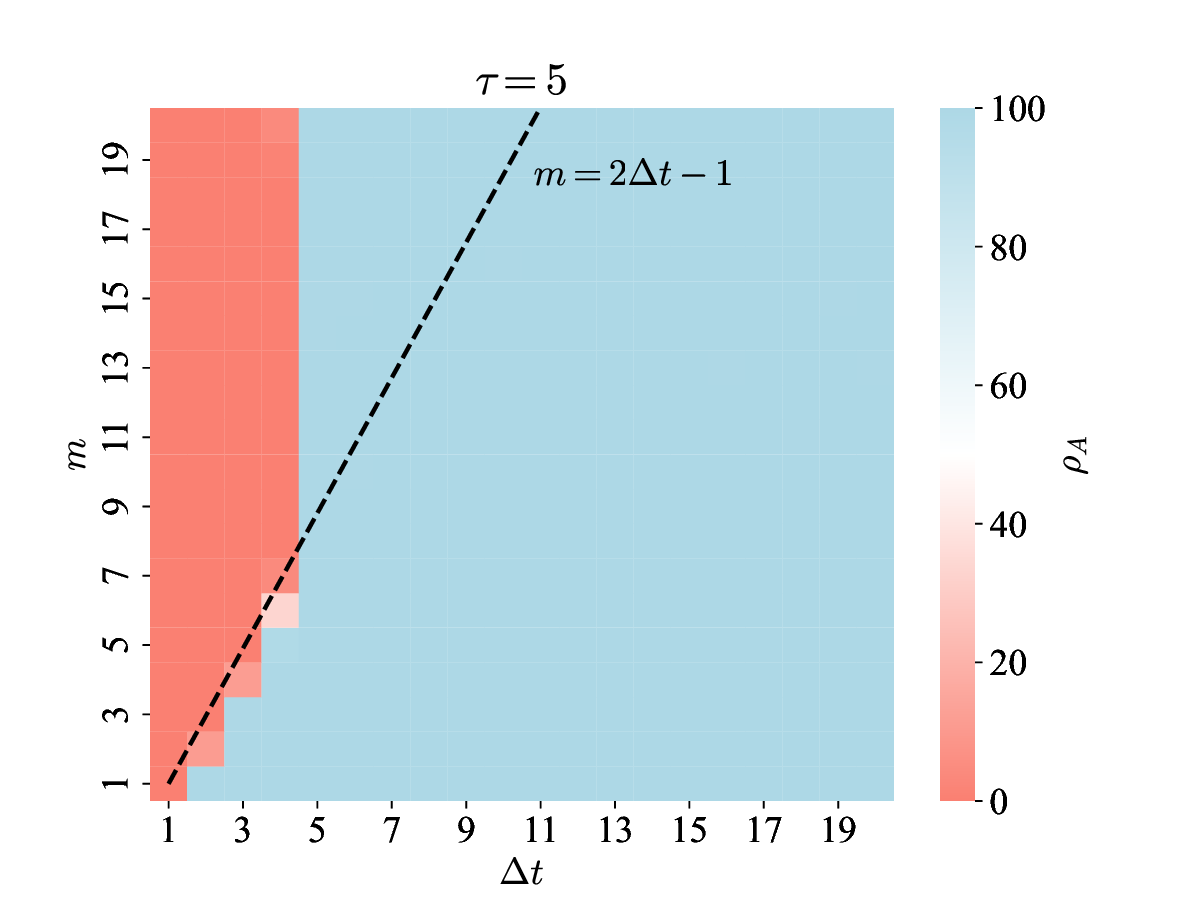}
\includegraphics[width=0.49\linewidth]{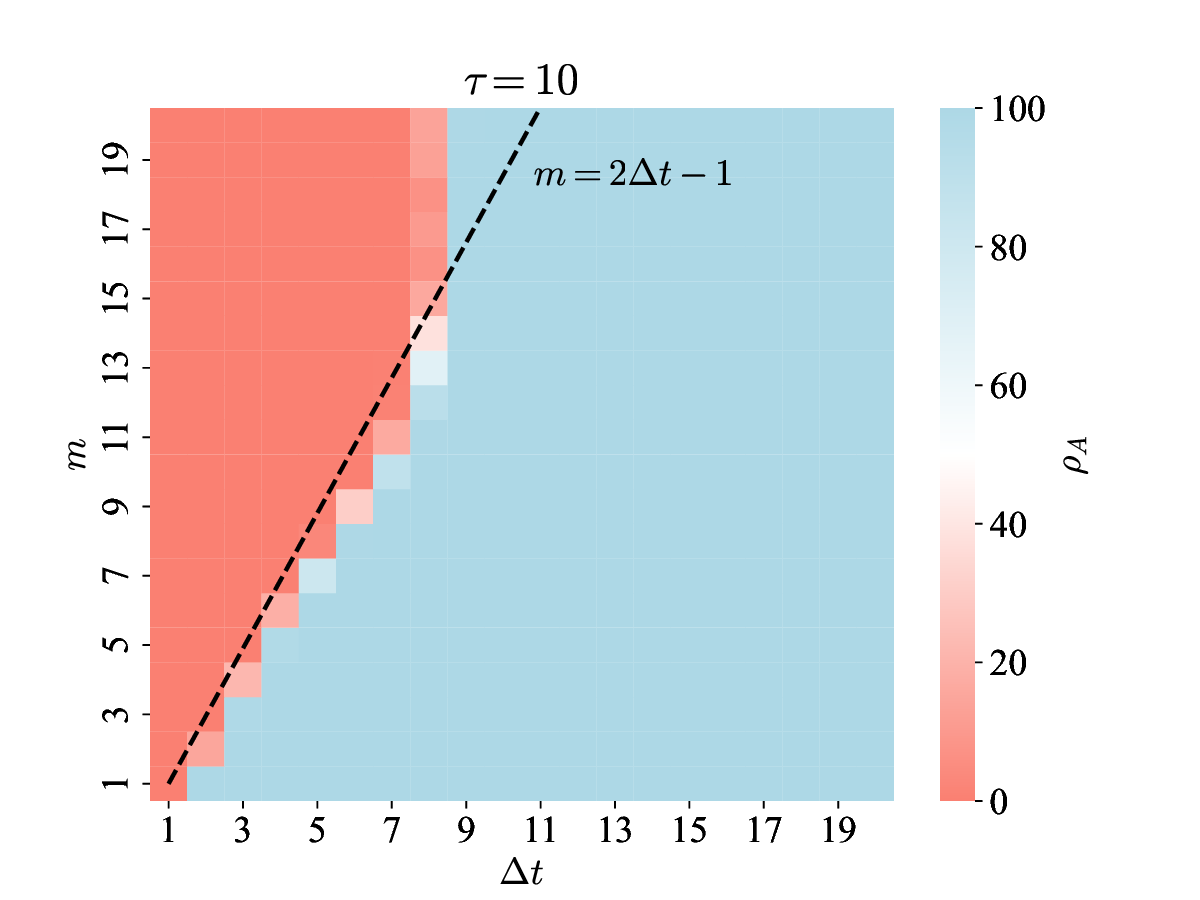}
\includegraphics[width=0.49\linewidth]{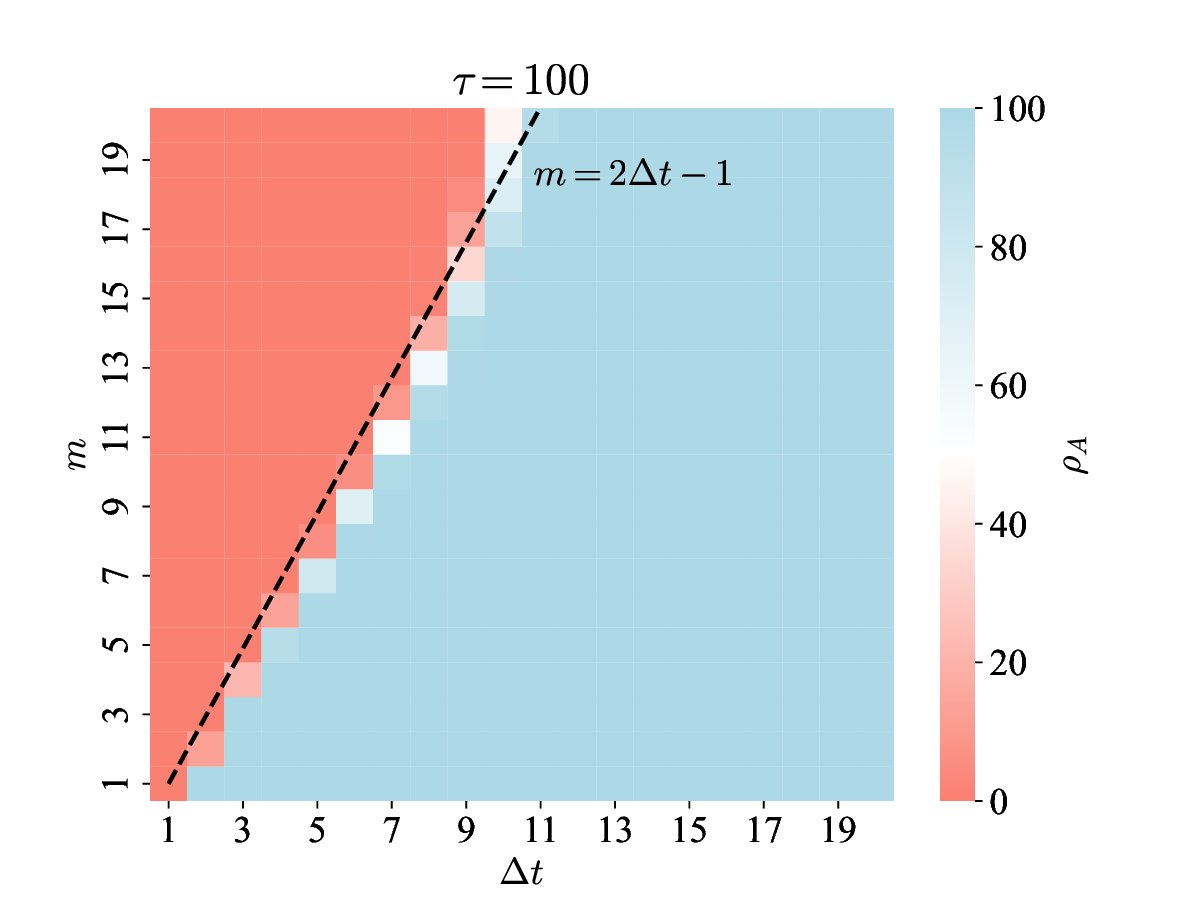}
\caption{Victory percentage of candidate $A$, $\rho_A$, shown as heatmaps in the $m \times \Delta t$ plane with $p_s=0.3$, for different values of the election horizon $\tau$. For $\tau = 4, 5, 10$ and $100$, the dotted line $m=2\Delta t -1$ represents the approximate boundary separating the 2 regions. Parameters: $N=1000$, $f_{A,0}=0.1$. results averaged over $r=500$ independent runs.}
\label{fig6}
\end{figure}

Figure \ref{fig6} shows the winning percentage of candidate $A$, denoted $\rho_A$, in the $m \times \Delta t$ plane for different values of the election horizon $\tau$, with $p_s=0.3$. This quantity is computed across all independent simulation runs. Specifically, at time $\tau$ we check in each run whether $f_A > 0.5$; if so, candidate A is recorded as the winner for that run at that time. Aggregating over all simulations yields the total number of victories for candidate $A$, from which $\rho_A$ is directly obtained. This procedure enables us to map how memory length and shock duration jointly affect candidate $A$'s likelihood of victory at different short-term horizons.

In addition, figure \ref{fig6} presents the evolution of $\rho_A$ as a function of the election horizon $\tau$. We begin by highlighting the general features of these results. For $\tau=1$, no regions of victory for candidate $A$ are observed, since $\rho_A=0$ across the entire plane. For $\tau \leq 2$, however, regions of victory start to emerge. At $\tau=2$, $\rho_A$ attains high values for $m=1$ over all $\Delta t \neq 1$, but decreases to zero for $m=4$, regardless of $\Delta t$. This behavior arises because low memory values rapidly freeze the dynamics after the shock event, thereby favoring candidate $A$, as also illustrated in the main panel of Figure \ref{fig4}. For larger values of $\tau$, the regions of victory evolve further, revealing how the interplay between memory and shock duration shapes candidate $A$'s chances over extended short-term horizons.

At $\tau=3$, the victory region of candidate $A$ expands considerably. The area initially located at $m=1$ for $\tau=2$ grows and spreads across the lower portion of the plane ($m<6$), while a new region emerges in the upper portion ($m>17$), both for $\Delta t>3$. The latter, however, corresponds only to a transient effect at low values of $\Delta t$. By contrast, the central region ($6<m<17$, also for $\Delta t>3$) remains close to the threshold, with $\langle f_A\rangle \approx 0.5$. For longer horizons, such as $\tau=5$ these patterns become even more pronounced, allowing us to assess whether candidate $A$'s advantage consolidates or fades as the system evolves further beyond the shock event

At $\tau=4$, the intermediate region also shifts in favor of candidate $A$, resulting in the largest overall victory area for this candidate. This implies that the most favorable scenario for candidate $A$ occurs when the election takes place four days after $t_0$, the onset of the electoral shock.

In addition, the heatmaps for $\tau = 4, 5, 10,$ and $100$ reveal that a dotted line given by $m = 2\Delta t - 1$ approximately separates the two opposing regions. The victory region of candidate $B$ expands continuously from $\tau=1$, but it always remains bounded by this line. Consequently, for candidate $A$ to maintain a high probability of victory at $\tau \gg 0$, the condition $m < 2\Delta t - 1$ must be satisfied.

This condition can be understood through a simple heuristic argument. During the shock interval of duration $\Delta t$, voters holding opinion $-1$ are repeatedly exposed to the external field favoring candidate $A$, which progressively imprints $+1$ states into their memory. Since each agent stores the last $m$ opinions, the shock effectively contributes of order $\Delta t$ favorable entries to the memory sequence. For the shock to induce a lasting realignment after it ends, these newly imprinted states must overcome the pre-shock memory content, which is dominated by opinion $-1$. This competition suggests that the memory length must not exceed a value proportional to the shock duration, leading naturally to a condition of the form $m \lesssim 2\Delta t$, in agreement with the numerically observed threshold $m < 2\Delta t - 1$. Although approximate, this argument provides a theoretical rationale for the scaling behavior identified in the simulations.

\begin{figure}
\centering
\includegraphics[width=0.7\linewidth]{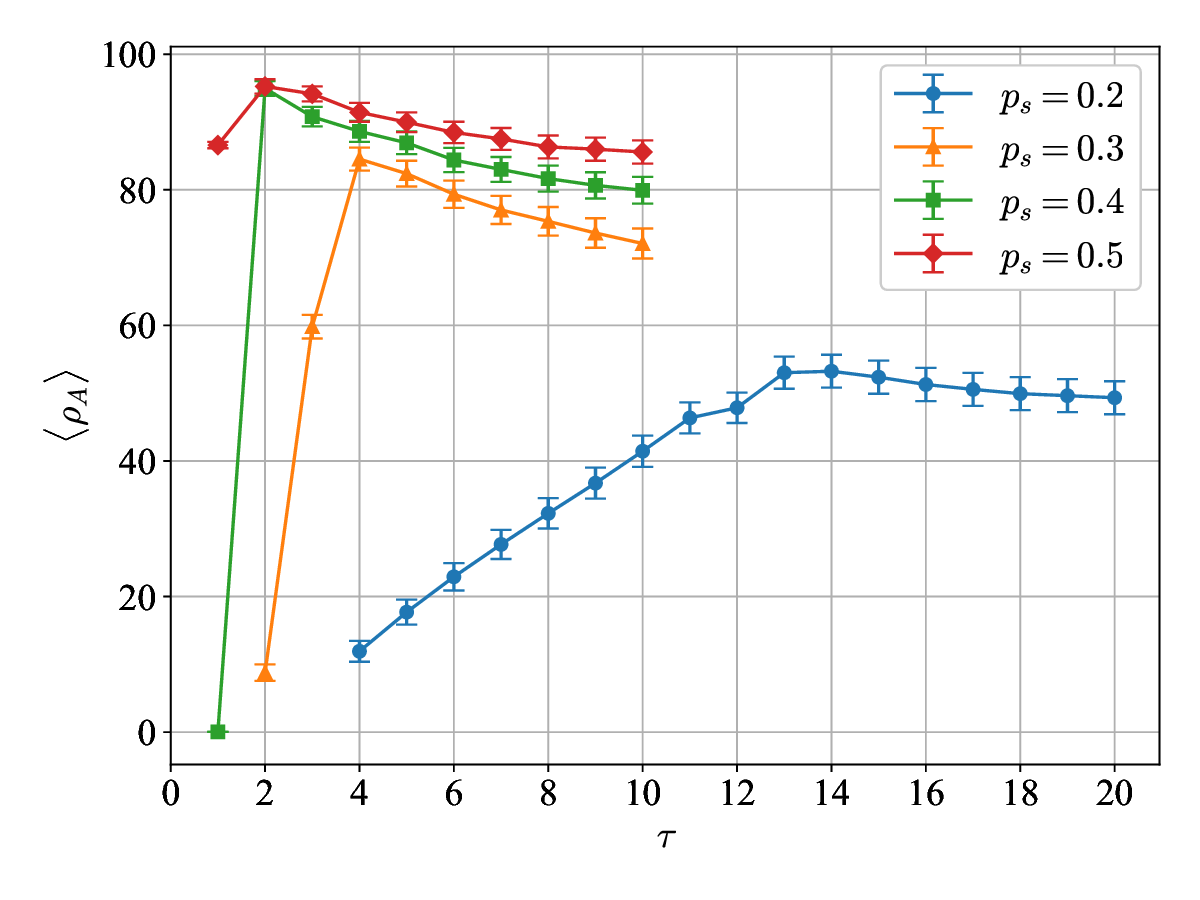}
\caption{Dependence of the average victory percentage of candidate $A$, $\langle \rho_A\rangle$, on the election horizon $\tau$, computed over the entire $m\times\Delta t$ plane, for several values of $p_s$. Parameters: $N=1000$, $f_{A,0}=0.1$, results averaged over $r=500$ independent simulations.}
\label{fig7}
\end{figure}

To complete our analysis of the most favorable scenario for candidate $A$, Figure \ref{fig7} presents the values of $\langle \rho_A \rangle$ for different $p_s$, where the averages were computed over the $m \times \Delta t$ plane (within the same range shown in Figure \ref{fig6}). This representation allows us to identify the optimal conditions for candidate $A$'s victory under various shock intensities. For higher values of $p_s$, smaller values of $\tau$ are more advantageous for candidate $A$ (e.g., $\tau=2$ for $p_s=0.4$ and $0.5$). Conversely, for lower values of $p_s$, longer times are required to reach the most favorable scenario (e.g., for $p_s=0.2$, the optimal value is $\tau=14$).

It is also important to note that the “best” scenario does not always correspond to a decisive advantage. For instance, when $p_s=0.2$, the most favorable case yields $\langle \rho_A \rangle \approx 53.24\% \pm 2.43\%$, which is only slightly above the majority threshold. By contrast, for $p_s=0.3$, $0.4$, and $0.5$, we obtain $\langle \rho_A \rangle > 84.5\%$, indicating a clear dominance of candidate $A$.

Taken together, these results show that the short-term impact of electoral shocks depends critically on their intensity, duration, and the memory length of voters. While strong shocks can secure an early and decisive advantage for candidate $A$, weaker shocks require longer time horizons and may only yield marginal gains, particularly when voter memory counteracts the transient effect of the external field.


\subsection{Qualitative comparison with real electoral shocks}

\qquad Although the present model is not intended to provide a quantitative fit to specific elections, it is instructive to compare its qualitative predictions with real electoral shocks observed in recent political processes. In particular, abrupt and exogenous events occurring during electoral campaigns offer empirical illustrations of the mechanisms explored here, namely the competition between voter inertia and sudden external perturbations.

A first illustrative example is the Brazilian presidential election of 2014, following the sudden death of candidate Eduardo Campos in August of that year. This unexpected event produced an immediate reconfiguration of voting intentions, with a rapid rise in support for his replacement, Marina Silva, as reported by national opinion polls in the days following the accident \cite{poll_marina1,poll_marina2}. However, this surge was not fully consolidated and gradually weakened as the election progressed, suggesting that the initial shock was insufficient to overcome longer-term preference inertia and the remaining time before the election. This behavior is qualitatively consistent with our model in regimes where the shock duration or intensity is moderate and the effective election horizon is large, leading to transient deviations that eventually fade. In terms of the model parameters, this situation corresponds to a large election horizon $\tau$ combined with a moderate shock intensity $p_s$.

A contrasting case is provided by the Brazilian presidential election of 2018, after the stabbing attack suffered by candidate Jair Bolsonaro in September. Polling data released shortly after the event indicated a noticeable increase in his voting intentions, exceeding typical statistical fluctuations \cite{poll1,poll2,poll3}. In this case, the shock occurred relatively close to the election day and was both intense and widely covered by the media, contributing to a rapid consolidation of electoral support. Within the framework of our model, this situation corresponds to a strong shock combined with a short election horizon, a regime in which the external perturbation can dominate voter memory and produce lasting realignments. Within the framework of the model, this situation is associated with a small election horizon $\tau$ and a strong shock intensity $p_s$.

These two examples, while not constituting a quantitative validation, qualitatively support the core phenomenology captured by the model: abrupt external shocks can temporarily or permanently reshape electoral outcomes, depending on their intensity, duration, and timing relative to the election. The comparison reinforces the interpretation of voter memory as a stabilizing mechanism and electoral shocks as destabilizing forces, whose interplay determines whether opinion shifts persist or dissipate.


\section{Final Remarks}
\label{conclusions}

\qquad In this work, we proposed a voter-like model for electoral competitions that integrates two fundamental mechanisms: individual memory and electoral shocks. By endowing voters with a finite memory of past opinions, we introduced persistence and inertia that slow down the dynamics and stabilize pre-existing preferences. At the same time, the model incorporates abrupt external influences through shocks of finite duration and intensity, representing real-world events such as scandals, impactful speeches, or breaking news.

Our simulations revealed that the two mechanisms act in competition: memory tends to preserve the status quo, while shocks can trigger sudden realignments. For weak shocks or long memory lengths, the system exhibits resilience, with transient deviations rapidly vanishing and the pre-shock configuration being restored. Conversely, shocks of sufficient strength or duration may decisively alter the outcome, even when candidate A initially faces highly unfavorable conditions. In particular, we found that the effectiveness of shocks depends sensitively on both their parameters and their timing, giving rise to phase-like regions in the $(p_s, \Delta t, m)$ space where either stability or realignment dominates.

Beyond reproducing intuitive features of electoral dynamics, our results highlight how short-term volatility and long-term inertia coexist in shaping collective decisions. This provides a quantitative framework for understanding why some extraordinary events dramatically alter election outcomes, while others fade without lasting impact. The model also suggests that resilience increases when voters rely more strongly on their personal history of opinions, underscoring the role of memory in buffering societies against abrupt perturbations.

Several extensions can be envisioned. A natural direction is to relax the assumption of a fully connected network by incorporating heterogeneous structures such as small-world or scale-free topologies, which are known to better capture social connectivity. Another possibility is to consider shocks favoring both candidates, or sequences of shocks, in order to mimic more complex campaign dynamics. Future work could also explore alternative formulations of memory and inertia—for instance, by varying the weighting of past opinions, introducing heterogeneous memory spans across individuals, or coupling memory to confidence-like variables. Such modifications may uncover richer dynamical regimes and provide closer parallels with cognitive and behavioral processes observed in real societies. Finally, the model could be calibrated with empirical electoral data to assess its predictive capacity in real scenarios.

Overall, the present framework contributes to the growing literature at the interface of statistical physics and political science, offering new insights into the interplay between persistence and sudden perturbations in electoral opinion dynamics.


\section*{Acknowledgments}
Nuno Crokidakis acknowledges  partial financial  support  from  the  Brazilian scientific funding  agency  Conselho Nacional de Desenvolvimento Cient\'ifico e Tecnol\'ogico (CNPq, Grant 308643/2023-2).

\end{document}